\documentclass[prb,byrevtex,floatfix,twocolumn,superscriptaddress]{revtex4}
\usepackage{amsmath}
\usepackage{braket}
\usepackage{amssymb,graphicx,afterpage}

\usepackage{txfonts}
\usepackage{multirow}
\usepackage{color}

\begin{document}

\title{Simultaneously Magnetic- and Electric-dipole Active Spin Excitations Govern the Static Magnetoelectric Effect in Multiferroic Materials}

\author{D\'avid Szaller}
\affiliation{Department of Physics, Budapest University of
Technology and Economics, 1111 Budapest, Hungary}

\author{S\'andor Bord\'acs}
\affiliation{Quantum-Phase Electronics Center,
Department of Applied Physics, The University of Tokyo, Tokyo
113-8656, Japan}

\author{Vilmos Kocsis}
\affiliation{Department of Physics, Budapest University of
Technology and Economics, 1111 Budapest, Hungary}

\author{Toomas R{\~o}{\~o}m}
\affiliation{National Institute of Chemical Physics and Biophysics,
Akadeemia tee 23, 12618 Tallinn, Estonia}

\author{Urmas Nagel}
\affiliation{National Institute of Chemical Physics and Biophysics,
Akadeemia tee 23, 12618 Tallinn, Estonia}

\author{Istv\'an K\'ezsm\'arki}
\affiliation{Department of Physics, Budapest University of
Technology and Economics, 1111 Budapest, Hungary}
\affiliation{Condensed Matter Research Group of the
Hungarian Academy of Sciences, 1111 Budapest, Hungary}

\date{\today}

\begin{abstract}
We derive a sum rule to demonstrate that the static magnetoelectric
(ME) effect is governed by optical transitions that are
simultaneously excited via the electric and magnetic components of
light. By a systematic analysis of magnetic point groups, we show
that the ME sum rule is applicable to a broad variety of
non-centrosymmetric magnets including ME multiferroic compounds. Due
to the dynamical ME effect, the optical excitations in these
materials can exhibit directional dichroism, i.e. the absorption
coefficient can be different for counter-propagating light beams.
According to the ME sum rule, the magnitude of the linear ME effect
of a material is mainly determined by the directional dichroism of
its low-energy optical excitations. Application of the sum rule to
the multiferroic Ba$_2$CoGe$_2$O$_7$, Sr$_2$CoSi$_2$O$_7$ and
Ca$_2$CoSi$_2$O$_7$ shows that in these compounds the static ME
effect is mostly governed by the directional dichroism of the
spin-wave excitations in the GHz-THz spectral range. On this basis,
we argue that the studies of directional dichroism and the
application of ME sum rule can promote the synthesis of new
materials with large static ME effect.
\end{abstract}

\maketitle

\section{Introduction}
Magnetoelectric (ME) multiferroics, where ferroelectricity coexists
with (ferro)magnetism, represent the most extensively studied class
of
multiferroics.\cite{FreemanSchmid,Fiebig2005,Eerenstein2006,Ramesh2007,Martin2010,Wu2013}
A spectacular control of the ferroelectric polarization by magnetic
field and manipulation of the magnetic order via electric field can
be realized in most of these materials as a direct consequence of
the coupling between spins and local electric dipoles. This offers a
fundamentally new path for data storage by combining the best
qualities of ferroelectric and magnetoresistive memories: fast
low-power electrical write operation, and non-destructive
non-volatile magnetic read operation.\cite{Martin2010,Wu2013} The
efficiency of multiferroics in such memory applications depends on
the strength of the magnetization-polarization coupling responsible
for the ME phenomena.

The ME effect has also been proposed to open new perspectives in
photonics. The entanglement between spins and local polarization
governs not only the ground-state properties but also the character
of excited states. Consequently, the electric component of light can
induce precession of the spins and the magnetic component of light
can generate electric polarization waves. This is termed as the
optical ME effect and has recently been observed for the spin
excitations in several multiferroic
compounds.\cite{Pimenov2006,Sushkov2007,Rovillain2010,Kezsmarki2011,
Bordacs2012,Ca2CoSi2O7,EuYMnO,Takahashi2013,Shuvaev2013}

As one of the most peculiar manifestations of the ME effect in the
optical regime, counter-propagating light beams can experience
different refractive indices in multiferroics. Strong directional
dichroism, that is difference in the absorption coefficient for
light beams traveling in opposite directions, has been reported for
spin excitations in these materials and proposed as a new principle
to design directional light switches operating in the GHz-THz
region.\cite{Kezsmarki2011, Bordacs2012, EuYMnO, Takahashi2013,
Ca2CoSi2O7}

Here, we show that optical studies of low energy magnons and phonons
in ME multiferroics, provide an efficient tool to further elucidate
microscopic mechanisms of multiferroicity. These studies can be
particularly useful to promote the systematic synthesis of new
materials with large static ME effect. We derive a relation,
hereafter referred to as the \emph{ME sum rule}, which shows the
connection between the static ME effect and the directional
dichroism observed for low-energy excitations. We specify the class
of materials where this ME sum rule is directly applicable. Finally,
we investigate the consequences of the ME sum rule for three
multiferroic materials, Ba$_2$CoGe$_2$O$_7$ (BCGO),
Sr$_2$CoSi$_2$O$_7$ (SCSO) and Ca$_2$CoSi$_2$O$_7$ (CCSO). For this
purpose, we compare their directional dichroism spectra to the
corresponding static ME coefficients reported in the
literature.\cite{Murakawa,Murakawa2,AkakiCCSO,Kusaka_CCSO,Sr2CoSi2O7_lowtemp,Sr2CoSi2O7_prb}
Absorption measurements used to determine the directional dichroism
in the GHz-THz spectral range were performed in the present study
and partly reproduced from our former
works.\cite{Kezsmarki2011,Bordacs2012,Ca2CoSi2O7}

The Kramers-Kronig relation, also known as the Hilbert
transformation, connects the real ($\Re$) and imaginary ($\Im$)
parts of a general frequency dependent response function
(susceptibility), $\chi(\omega)$, which corresponds to a linear and
causal response function in the time domain:
\begin{eqnarray*}
\Re\chi(\omega)&=&\frac{1}{\pi}\mathcal{P}\int^{\infty}_{-\infty}{\frac{\Im\chi(\omega^{\prime})}{\omega^{\prime}-\omega}\textrm{d}\omega^{\prime}},\\
\Im\chi(\omega)&=&-\frac{1}{\pi}\mathcal{P}\int^{\infty}_{-\infty}{\frac{\Re\chi(\omega^{\prime})}{\omega^{\prime}-\omega}\textrm{d}\omega^{\prime}},
\label{KK_frequency}
\end{eqnarray*}
where $\mathcal{P}$ stands for the Cauchy principal value integral.
In many cases, either the real or the imaginary part of
$\chi(\omega)$ can be determined experimentally and the
Kramers-Kronig transformation is used to obtain the entire complex
response function. In the limit of $\omega$$=$$0$, these expressions
are simplified to the following form, which shows close similarity
with sum rules:
\begin{eqnarray}
\Re\chi(\omega=0)&\equiv&\chi(0)=\frac{2}{\pi}\mathcal{P}\int^{\infty}_{0}{\frac{\Im\chi(\omega)}{\omega}\textrm{d}\omega}, \label{KK_static_Re}\\
\Im\chi(\omega=0)&\equiv&0=-\frac{1}{\pi}\mathcal{P}\int^{\infty}_{-\infty}{\frac{\Re\chi(\omega)}{\omega}\textrm{d}\omega}.
\label{KK_static_Im}
\end{eqnarray}
Equation \ref{KK_static_Re} shows that the static response of a
system is fully determined by the corresponding dynamical
susceptibility and the frequency denominator on the right-hand side
indicates the vital role in low-energy excitations to the static
susceptibility.

A common example is the dielectric permittivity of semiconductors,
which is usually larger for compounds with smaller charge gap and
can be considerably affected by the contributions from low-energy
phonon modes. A particularly strong enhancement is found in quantum
paraelectrics due to the presence of soft polar phonon
modes.\cite{SrTiO3_1969, EuTiO3_2007} Besides low-energy or soft
modes, in materials with ferroic orders, the ac susceptibility
related to the domain dynamics can also influence the static
response.

In multiferroic materials the coupling between the electric
polarization and the magnetization can be phenomenologically
described by the magnetoelectric susceptibility tensors
$\chi^{me}(\omega)$ and $\chi^{em}(\omega)$, where $\Delta
M_{\gamma}^{\omega}=\chi^{me}_{\gamma\delta}(\omega)E_{\delta}^{\omega}$
is the magnetization generated by an oscillating electric field and
$\Delta
P_{\delta}^{\omega}=\chi^{em}_{\delta\gamma}(\omega)H_{\gamma}^{\omega}$
is the polarization induced by an oscillating magnetic field,
respectively. Here $\gamma$ and $\delta$ stand for the Cartesian
coordinates and the two cross-coupling tensors are connected by the
$\{\ldots\}^{\prime}$ time-reversal operation according to
$\{\chi^{me}_{\gamma\delta}(\omega)\}^{\prime}=-\chi^{em}_{\delta\gamma}(\omega)$.

In a broad class of materials lacking simultaneously spatial
inversion and time reversal symmetries,\cite{Hornreich1968,
Dell1970, Arima2008, Cano2009, DDPRB2013} including also
multiferroic compounds, the time reversal odd part of the ME
susceptibility
can induce a difference in the complex refractive index of
counter-propagating electromagnetic waves,
\begin{equation}
\begin{split}
N^{\pm}(\omega)\approx\sqrt{\varepsilon_{\delta\delta}(\omega)\mu_{\gamma\gamma}(\omega)}\pm\frac{1}{2}\left[\chi^{me}_{\gamma\delta}(\omega)-\{\chi^{me}_{\gamma\delta}(\omega)\}^{\prime}\right].
\label{N_pm}
\end{split}
\end{equation}
Here $N^{\pm}$ stands for the refractive indices of waves
propagating in opposite directions $(\pm \boldsymbol{k})$. The
$\boldsymbol{e}_\delta$ and $\boldsymbol{e}_\gamma$ unit vectors are
parallel to the direction of the electric ({\bf E}$^{\omega}$) and
magnetic ({\bf H}$^{\omega}$) fields of light, respectively, while
 $\varepsilon_{\delta\delta}(\omega)$ and $\mu_{\gamma\gamma}(\omega)$ are diagonal components of the complex relative permittivity and permeability tensors in the $\{\boldsymbol{e}_\delta, \boldsymbol{e}_\gamma, \boldsymbol{e}_\eta$$\parallel$$\boldsymbol{k}\}$ basis.
From this point on we restrict our study to those cases, when the
solutions of the Maxwell equations are linearly polarized waves or
the linear polarization of the incident light is nearly preserved
during the propagation through the magnetoelectric medium. This
condition needs to be satisfied to have direct comparison between
the static and optical ME data. The difference in the imaginary part
of the $N^{+}$ and $N^{-}$ refractive indices gives rise to a
difference in the absorption coefficients of counter-propagating
waves, termed as directional dichroism:
\begin{equation}
\begin{split}
\Delta\alpha(\omega)=\alpha_{+}(\omega)-\alpha_{-}(\omega)=\frac{2\omega}{c}\Im(\chi^{me}_{\gamma\delta}(\omega)-\{\chi^{me}_{\gamma\delta}(\omega)\}^{\prime}),
\label{dalpha_pm}
\end{split}
\end{equation}
where $c$ is the speed of light in vacuum.

\section{Results}
\subsection{The ME sum rule}
In several classes of non-centrosymmetric magnets
$\chi^{me}_{\gamma\delta}(\omega)$ is antisymmetric with respect to
the time reversal, as listed in Table \ref{table1} and discussed
later in this article. In this case, the static ME properties and
the optical directional dichroism are described by the same element
of the ME tensor, hence, Eqs.~1 and 4 yield the following ME sum
rule:
\begin{equation}
\begin{split}
\chi^{me}_{\gamma\delta}(0)=\frac{c}{2\pi}\mathcal{P}\int^{\infty}_{0}{\frac{\Delta\alpha(\omega)}{\omega^{2}}\textrm{d}\omega}.
\label{KhiME_DC}
\end{split}
\end{equation}
According to this sum rule the static ME effect is mostly governed
by the directional dichroism of low-energy excitations, since the
absorption difference, $\Delta\alpha$, is cut off by the
$\omega^{2}$ denominator at higher frequencies.
The ME sum rule in Eq. \ref{KhiME_DC} can also be derived using the
Kubo formula as described in the Appendix.

Following Neumann's principle, we specify the symmetry of those
magnetic crystals for which $\chi^{me}_{\gamma\delta}(\omega)$
changes sign upon the time reversal.
This off-diagonal ME tensor component is antisymmetric if and only
if
\begin{itemize}
\item[(1)] all spatial symmetry operations of the
magnetic point group (MPG) transforming
$\chi^{me}_{\gamma\delta}(\omega)$ into
$-\chi^{me}_{\gamma\delta}(\omega)$ are combined with the time
reversal operation, and there is at least one such symmetry
operation present in the MPG and
\item[(2)] none of the spatial symmetry operations that leave
$\chi^{me}_{\gamma\delta}(\omega)$ invariant are combined with time
reversal and
\item[(3)] symmetry elements connecting
$\chi^{me}_{\gamma\delta}(\omega)$ to
$\chi^{me}_{\delta\gamma}(\omega)$ or
$-\{\chi^{me}_{\delta\gamma}(\omega)\}^\prime$ and symmetry elements
transforming $\chi^{me}_{\gamma\delta}(\omega)$ to
$-\chi^{me}_{\delta\gamma}(\omega)$ or
$\{\chi^{me}_{\delta\gamma}(\omega)\}^\prime$ are not present in the
MPG at the same time.
\end{itemize}
\begin{table*}[!t]
\caption{Crystallographic magnetic point groups (MPG) hosting
$\chi^{me}_{xy}$ and $\chi^{me}_{xz}$ ME tensor elements, which are
antisymmetric with respect to the time reversal, listed in the
second and fifth columns, respectively. Here $z$ denotes the
principal symmetry axis and MPGs are labeled in the international
notation. The subscripts of the symmetry operations show the axes of
the $n$ proper and $\overline{n}$ improper rotations and the axes
perpendicular to the $m$ mirror planes. Subscript $d$ denotes the
diagonal direction between the $x$ and $y$ coordinate axes. Symmetry
operations marked by prime ($\prime$) are combined with the time
reversal. The $\chi^{me}_{xy}$ and $\chi^{me}_{xz}$ ME tensor
elements correspond to light propagation along the $z$ and $y$ axes,
respectively. For the MPGs marked with asterisks in the third and
sixth columns, the solutions of the Maxwell equations in the
transverse-wave approximation are linearly polarized waves. The few
remaining MPGs are chiral, hence, they show circular dichroism.
Several example materials are given in the fourth and seventh
columns, where $H_\alpha$ --if specified-- stands for an external
magnetic field pointing to the $\alpha$ crystallographic direction.
In these cases $x$ and $z$ are the actual high-symmetry axes, i.e.
for $H_{[100]}$, $H_{[001]}$ and $H_{[110]}$ the corresponding
coordinates are $x$, $z$ and again $x$, respectively. In hexagonal
manganites ScMnO$_3$ and LuMnO$_3$, there are coexisting magnetic
phases with sample dependent temperature ranges,\cite{RMnO3} thus,
they are indicated in two lines of the table.}
\begin{tabular}{ccccccc}
\hline \hline
 Crystal & \multicolumn{2}{c}{\multirow{2}{*}{$\chi^{me}_{xy}(\omega)=-\{\chi^{me}_{xy}(\omega)\}^{\prime}$}} & \multirow{2}{*}{Materials} & \multicolumn{2}{c}{\multirow{2}{*}{$\chi^{me}_{xz}(\omega)=-\{\chi^{me}_{xz}(\omega)\}^{\prime}$}} & \multirow{2}{*}{Materials} \\
 system &  &  &  & \\
\hline
 Triclinic & $\overline{1}^{\prime}_{z}$ & * & & $\overline{1}^{\prime}_{z}$ & * & \\
\hline
 \multirow{2}{*}{Monoclinic} & $m^{\prime}_{z}$ & * & Ni$_3$B$_7$O$_{13}$I;\cite{Ni3B7O13I} BiTeI $H_{[100]}$\cite{BiTeI} & $2^{\prime}_{z}$ &  &  LiCoPO$_4$;\cite{LiCoPO4} Cu$_2$OSeO$_3$ $H_{[110]}$;\cite{Cu2OSeO3} Co$_3$TeO$_6$\cite{Co3TeO6}\textsuperscript{,}\footnote{NdFe$_3$(BO$_3$)$_4$  $H_{[010]}$\cite{NdFeborat}}\\
  & $2_{z}m^{\prime}_{z}$ & * & TbOOH;\cite{TbOOH} Ba$_2$Ni$_3$F$_{10}$\cite{Ba2Ni7F18andCsCoF4} & $2^{\prime}_{z}m_{z}$ & * &  TbPO$_4$;\cite{TbPO4} MnPS$_3$;\cite{MnPS3} Co$_3$TeO$_6$ 17 K $<T<$ 21 K\cite{Co3TeO6}\\
\hline
 \multirow{2}{*}{Rhombic}  & $m_{x}m_{y}m^{\prime}_{z}$ & * & LiNiPO$_4$\cite{LiNiPO4}\textsuperscript{,}\footnote{one-dimensional photonic crystal with four-layered unit cell\cite{Figotin_prE}} & $m_{x}m^{\prime}_{y}2^{\prime}_{z}$ & * &  BCGO,\cite{Kezsmarki2011} SCSO,\cite{Sr2CoSi2O7_prb,Sr2CoSi2O7_lowtemp} CCSO,\cite{Ca2CoSi2O7} CuB$_2$O$_4$\cite{SaitoJPSJ2008} $H_{[110]}$\footnote{CdS,\cite{Hopfield} AlN, GaN, InN $H_{[100]}$;\cite{wurtzite} CaBaCo$_4$O$_7$;\cite{CaBaCo4O7} GaFeO$_3$;\cite{Kubota2004}  Co$_3$B$_7$O$_{13}$Br;\cite{CoB7O13Br} KMnFeF$_6$\cite{Ba2Ni7F18andCsCoF4}}\\
  &  &  &  & $2^{\prime}_{x}2_{y}2^{\prime}_{z}$ &  & BCGO,\cite{Bordacs2012} SCSO,\cite{Sr2CoSi2O7_prb,Sr2CoSi2O7_lowtemp} CCSO,\cite{Ca2CoSi2O7} CuB$_2$O$_4$\cite{SaitoPRL2008} $H_{[100]}$\footnote{Cu$_2$OSeO$_3$ $H_{[100]}$;\cite{Cu2OSeO3} [Ru(bpy)$_2$(ppy)][MnCr(ox)]
;\cite{oxalate} one-dimensional photonic crystal with three-layered unit cell\cite{Figotin_prE}} \\
\hline
 \multirow{4}{*}{Tetragonal} & $4_{z}m^{\prime}_{z}$ & * &  & $$ & &  \\
  & $2^{\prime}_{x}m_{d}\overline{4}^{\prime}_{z}$ & * &  & $$ & &  \\
  & $2^{\prime}_{x}2^{\prime}_{d}4_{z}$ &  & Nd$_5$Si$_4$\cite{Nd5Si4} & $$ & &  \\
  & $m_{x}m_{d}4_{z}m^{\prime}_{z}$ & * &  & $$ & &  \\
\hline
 \multirow{5}{*}{Rhombohedral} & $\overline{3}^{\prime}_{z}$ & * & Cr$_2$O$_3$\cite{Cr2O3} & $\overline{3}^{\prime}_{z}$ & * &  Cr$_2$O$_3$\cite{Cr2O3} \\
  &  &  &  & $m^{\prime}_{y}3_{z}$ & * & BiTeI $H_{[001]}$\cite{BiTeI} \\
  & $2^{\prime}_{x}3_{z}$ &  &  & $2^{\prime}_{x}3_{z}$ &  &  \\
  & $m_{x}\overline{3}^{\prime}_{z}$ & * & Gd$_2$Ti$_2$O$_7$\cite{Gd2Ti2O7} & $m_{x}\overline{3}^{\prime}_{z}$ & * & Gd$_2$Ti$_2$O$_7$\cite{Gd2Ti2O7} \\
  &  &  &  & $m^{\prime}_{y}\overline{3}^{\prime}_{z}$ & * & Nb$_2$Mn$_4$O$_9$, Nb$_2$Co$_4$O$_9$\cite{Nb2Mn4O9} \\
\hline
 \multirow{5}{*}{Hexagonal} & $\overline{6}^{\prime}_{z}$ & * &  & $6^{\prime}_{z}$ &  & ScMnO$_3$, LuMnO$_3$\cite{RMnO3} \\
  & $6_{z}m^{\prime}_{z}$ & * &  & $6^{\prime}_{z}m_{z}$ & * &  \\
  & $m_{x}2^{\prime}_{y}\overline{6}^{\prime}_{z}$ & * &  & $2^{\prime}_{x}m^{\prime}_{y}\overline{6}_{z}$ & * & Fe$_2$P\cite{Fe2P} \\
  & $2^{\prime}_{x}2^{\prime}_{y}6_{z}$ &  &  & $m_{x}m^{\prime}_{y}6^{\prime}_{z}$ & * & HoMnO$_3$;\cite{HoMnO3,RMnO3} YMnO$_3$, ErMnO$_3$, YbMnO$_3$\cite{RMnO3}\textsuperscript{,\footnote{ScMnO$_3$, LuMnO$_3$, TmMnO$_3$\cite{RMnO3}}} \\
  & $m_{x}m_{y}6_{z}m^{\prime}_{z}$ & * &  & $2_{y}6^{\prime}_{z}$ &  &  \\
  &  &  &  & $m_{x}m^{\prime}_{y}6^{\prime}_{z}m_{z}$ & * &  \\
\hline \hline
\end{tabular}
\label{table1}
\end{table*}

When light propagates along the principal axis of the crystal
labeled as the $z$ axis, the off-diagonal tensor component
$\chi^{me}_{xy}(\omega)$ can generate directional dichroism, where
$x$ and $y$ axis are perpendicular to the $z$ axis. These three
conditions are fulfilled for $\chi^{me}_{xy}(\omega)$ if the MPG
meets all of the following criteria: i) it contains either an
$\overline{n}^{\prime}_{z}$ or $2^{\prime}_{x}$ symmetry operation
where $n \in \left\{ 1,2,3,6 \right\}$, ii) the MPG does not have
any element combined with time reversal besides the previous ones
and $\overline{4}^{\prime}_{z}$, and iii) the MPG does not contain
any of the previously listed operations without a subsequent time
reversal. Here subscripts stand for the axis of the $n$-fold proper
$(n)$ or improper $(\overline{n})$ rotations, primes ($^\prime$)
following spatial transformations indicate the time reversal
operation.

For light propagation along the $y$ axis perpendicular to the
principal $z$ axis, $\chi^{me}_{xz}(\omega)$ can generate
directional dichroism. In this case, the conditions (1)-(3)
specifying the requirements of $\chi^{me}_{xz}(\omega)$ being
antisymmetric with respect to the time reversal are fulfilled if the
MPG matches all of the following criteria: i) it contains at least
one symmetry element from
$\left\{\overline{1}^{\prime},\overline{2}^{\prime}_{x},2^{\prime}_{y},2^{\prime}_{z},\overline{3}^{\prime}_{z},6^{\prime}_{z}\right\}$,
ii) the MPG does not have any element combined with the time
reversal besides the previous ones, and iii) the MPG does not
contain $4_{z}$, $\overline{4}_{z}$ or any of the previously listed
operations without a subsequent time reversal.

The MPGs fulfilling these requirements are listed in Table
\ref{table1} together with example materials. For MPGs marked by
asterisks, the refractive index of the corresponding materials is
described by Eq.~\ref{N_pm}. In all these MPGs, the solutions of the
Maxwell-equations are linearly polarized waves. In some of these
cases, when there is a finite magnetization perpendicular to the
light propagation, the polarization can have a small longitudinal
component, which is neglected here. This transverse-wave
approximation means that we neglect additional terms in the
refractive index, which are higher-order products of tensor
components like $\chi^{me}_{yz}\epsilon_{zx}/\epsilon_{zz}$ or
$\chi^{me}_{zy}\mu_{zx}/\mu_{zz}$ for propagation along the $z$
axis.

In materials belonging to MPGs not marked by asterisk, natural and
magnetic circular dichroism can appear, since these MPGs are all
chiral and in some cases finite magnetization is allowed parallel to
the light propagation direction (Faraday configuration). However,
for sufficiently thin samples the linear polarization of the
incident light is nearly preserved even then. Thus, the index of
refraction can be approximated by Eq.~\ref{N_pm} for all of the
listed MPGs. This allows a direct comparison between the static and
dynamical ME effects according to the ME sum rule in
Eq.~\ref{KhiME_DC}, since the static measurements used to determine
the off-diagonal ME tensor elements can be compared to the optical
experiments with linearly polarized light. In the second and third
rows of Fig.~\ref{fig1} the $2^{\prime}_{x}2_{y}2^{\prime}_{z}$
chiral state of BCGO\cite{Bordacs2012} and CCSO\cite{Ca2CoSi2O7} is
studied in the Faraday configuration, where the material shows
polarization rotation. Nevertheless, the directional dichroism can
be well approximated by Eq.~\ref{dalpha_pm}.\cite{Bordacs2012}

\subsection{Application of the ME sum rule to multiferroic materials}
In order to check the applicability of the ME sum rule, we compare
the magnetic field dependence of the static and optical ME effects
for three members of the multiferroic melilite family, namely for
Ba$_2$CoGe$_2$O$_7$, Ca$_2$CoSi$_2$O$_7$ and Sr$_2$CoSi$_2$O$_7$.
These compounds crystallize in the non-centrosymmetric tetragonal
P$\overline{4}$2$_1$m
structure\cite{BCGO_Zheludev,BCGO_Hutanu2011,CCSO_struct_Hagiya,CCSO_struct_Jia}
where Co$^{2+}$ cations with S=3/2 spin form square-lattice layers
stacked along the tetragonal [001] axis. They undergo an
antiferromagnetic transition at T$_N$$\approx$6-7\,K. Due to strong
single-ion anisotropy, the two-sublattice antiferromagnetic state
has an easy-plane character with spins lying within the tetragonal
plane.\cite{SatoPhysB,BCGO_Zheludev,BCGO_Hutanu2012,AkakiCCSO,ACSO}
The free rotation of the magnetization within the tetragonal plane
can already be realized by moderate fields of $\lesssim$1-2\,T,
which is an indication of a weak in-plane
anisotropy.\cite{Murakawa,Bibe_prb_new} As another consequence of
the single-ion anisotropy, the magnetization is saturated at
different magnetic field values, $H^{Sat}_{plane}$ and
$H^{Sat}_{axis}$, when the field is applied within the easy plane
and along the hard axis, respectively. Prior to saturation, the
magnetization follows a nearly linear field dependence due to the
increasing canting of the sublattice moments for any direction of
the magnetic field.

\begin{figure*}[t!]
\includegraphics[width=7.05in]{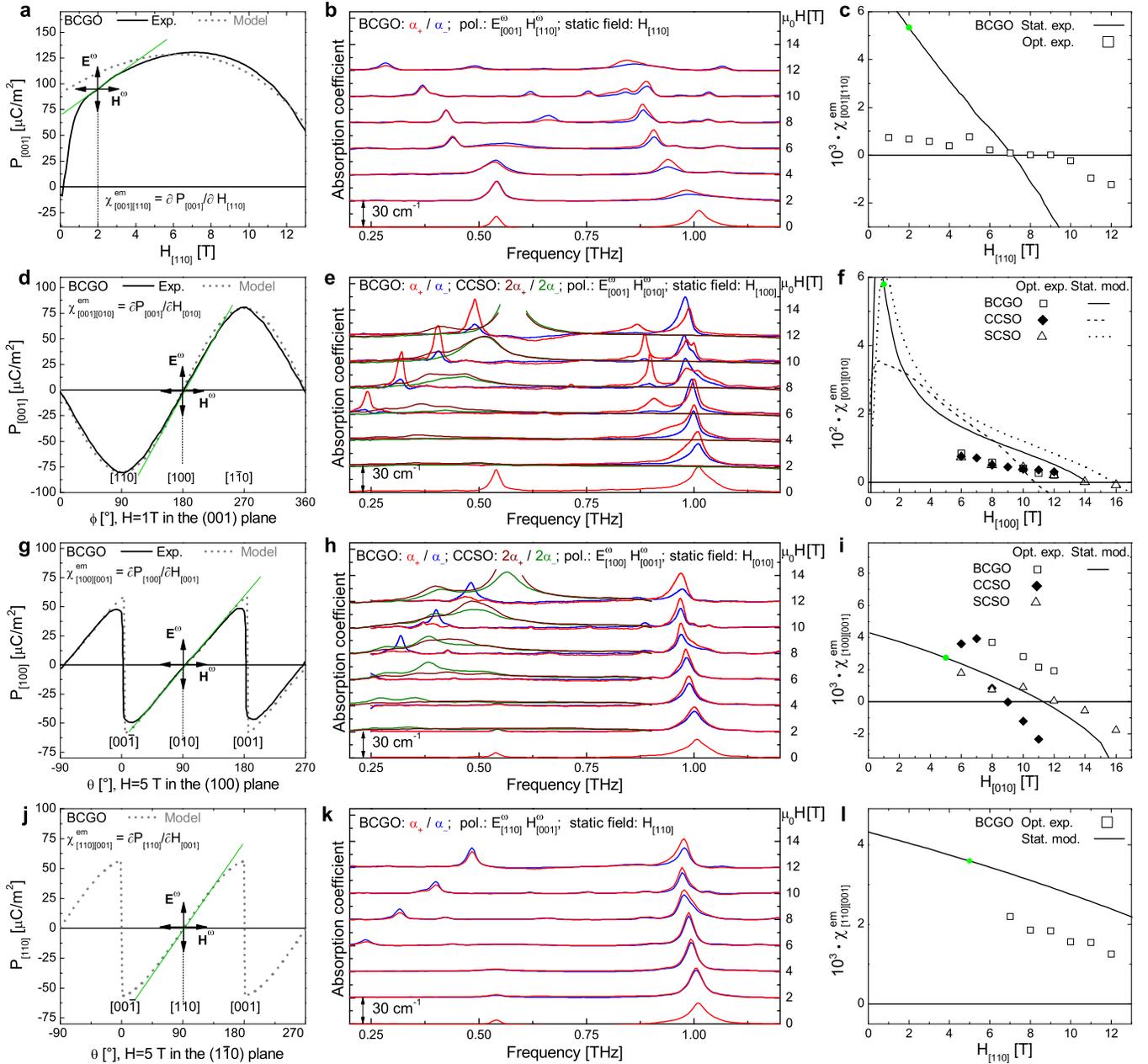}
\caption{(Color online) Comparison of the static and optical ME
properties of multiferroic Ba$_2$CoGe$_2$O$_7$ (BCGO),
Ca$_2$CoSi$_2$O$_7$ (CCSO) and Sr$_2$CoSi$_2$O$_7$ (SCSO) based on
the ME sum rule in Eq.~\ref{KhiME_DC}. Panel (a): Dependence of the
ferroelectric polarization ($P$) on the magnitude of the magnetic
field ($H$) in BCGO. Panels (d), (g) and (j): Dependence of $P$ on
the orientation of the field in BCGO. In these panels the solid
lines are experimental data reproduced from
Ref.~\onlinecite{Murakawa}, while the dashed lines are calculated
using the spin dependent p-d hybridization model according to
Eqs.~\ref{Pa}-\ref{Pc}.\cite{Murakawa} The slopes of the green lines
in the same panels are proportional to the corresponding elements of
the ME tensor. Arrows labeled with ${\bf E}^\omega$ and ${\bf
H}^\omega$ show the electric and magnetic components of the absorbed
light in the corresponding optical experiment, respectively. Panels
(b), (e), (h) and (k): Field dependence of the magnon absorption
spectra of BCGO in the GHz-THz range. The light polarizations
indicated in these panels correspond to the labels ${\bf E}^\omega$
and ${\bf H}^\omega$ shown in the panels of the first column. The
spectra are shifted vertically proportional to $H$. For BCGO, the
spectra corresponding to counter-propagating light beams are plotted
by red and blue lines, while for CCSO brown and dark green lines
represent the two propagation directions. The absorption coefficient
of CCSO is multiplied by a factor of two for better visibility. The
spectra in panels (b) and (k) are measured in the present study,
while the data in panels (e) and (h) are reproduced from
Ref.~\onlinecite{Bordacs2012} for BCGO and from
Ref.~\onlinecite{Ca2CoSi2O7} for CCSO, respectively. Panels (c),
(f), (i) and (l): Magnetic field dependence of different components
of the ME tensor. Symbols indicate the tensor elements calculated
from the corresponding optical measurements using the ME sum rule;
empty square, full diamond and empty triangle stand for BCGO, CCSO
and SCSO, respectively. The field dependence of the static ME tensor
components are plotted with solid, dashed and dotted lines for the
three compounds in the same order. The points corresponding to the
slope of the green lines in the left panels are indicated by a green
dot. The solid line in panel (c) is calculated directly from the
measured polarization-magnetic field curve shown on panel (a), while
the curves in panels (f), (i) and (l) are evaluated using
Eqs.~\ref{Pa}-\ref{Pc}. Static experiments, optical measurements and
model calculations were carried out at $T$$=$$2$\,K, $T$$=$$4$\,K
and $T$$=$$0$\,K, respectively.} \label{fig1}
\end{figure*}

The multiferroic character of these materials has been intensively
studied both theoretically,\cite{Picozzi2011, Perez_Mato2011,
Toledano2011} and experimentally\cite{Murakawa,Murakawa2,
YiBCGO,AkakiCCSO,ACSO,Sr2CoSi2O7_prb,Sr2CoSi2O7_lowtemp} via their
static ME properties. The strong optical ME effect emerging at their
spin-wave excitations has also attracted much
interest.\cite{Kezsmarki2011,Bordacs2012,Ca2CoSi2O7,Miyahara2011,Miyahara2012,Karlo_prl,Matsumoto2013}
The magnetically induced ferroelectric polarization has been
described by the spin-dependent hybridization of the Co$^{2+}$ $d$
orbitals with the $p$ orbitals of the surrounding oxygen ions
forming tetrahedral cages.\cite{Murakawa,Murakawa2} When the
magnetization is a linear function of the applied field, the
direction of the sublattice magnetizations can be straightforwardly
expressed as a function of the orientation and the magnitude of the
magnetic field. Then, the components of the magnetically induced
ferroelectric polarization are directly determined from the
orientation of the sublattice magnetizations within the
spin-dependent hybridization model:\cite{Arima_pd,Murakawa}
\begin{widetext}
\begin{eqnarray}
P_{[100]} & = & A_{plane}\left[\frac{H\textrm{sin}\theta}{H^{Sat}_{plane}}-\sqrt{1-\left(\frac{H\textrm{sin}\theta}{H^{Sat}_{plane}}\right)^2}\right]\sqrt{1-\left(\frac{H\textrm{cos}\theta}{H^{Sat}_{axis}}\right)^2}\frac{H\textrm{cos}\theta}{H^{Sat}_{axis}}\textrm{sin}\phi\label{Pa},\\
P_{[010]} & = & A_{plane}\left[\frac{H\textrm{sin}\theta}{H^{Sat}_{plane}}-\sqrt{1-\left(\frac{H\textrm{sin}\theta}{H^{Sat}_{plane}}\right)^2}\right]\sqrt{1-\left(\frac{H\textrm{cos}\theta}{H^{Sat}_{axis}}\right)^2}\frac{H\textrm{cos}\theta}{H^{Sat}_{axis}}\textrm{cos}\phi\label{Pb},\\
P_{[001]} & = &
A_{axis}\left[\left(\frac{H\textrm{sin}\theta}{H^{Sat}_{plane}}\right)^2-\frac{H\textrm{sin}\theta}{H^{Sat}_{plane}}\sqrt{1-\left(\frac{H\textrm{sin}\theta}{H^{Sat}_{plane}}\right)^2}-\frac{1}{2}\right]\left[1-\left(\frac{H\textrm{cos}\theta}{H^{Sat}_{axis}}\right)^2\right]\textrm{sin}2\phi\label{Pc}.
\end{eqnarray}
\end{widetext}
Here $\theta$ and $\phi$ are the polar and azimuthal angles of the
magnetic field relative to the $[001]$ and $[100]$ axes,
respectively, and $H$ is the magnitude of the field. $A_{plane}$ and
$A_{axis}$ are constants describing the strength of the
magnetoelectric coupling. To make the formulas more compact, the
tilting angle of the two inequivalent oxygen tetrahedra in the unit
cell was approximated by $\pi/4$, which is close to the experimental
value of $48^{\circ}$ for CCSO.\cite{Kusaka_CCSO} For BCGO, the
saturation fields are $H^{Sat}_{plane}$$\approx$$16$\,T and
$H^{Sat}_{axis}$$\approx$$36$\,T as found both in the
static\cite{Bibe_prb_new} and optical experiments.\cite{Karlo_prl}
By fitting the field dependence of the static polarization
reproduced from Ref.~\onlinecite{Murakawa,Murakawa2} in
Fig.~\ref{fig1}(a) and (g), we obtain $A_{plane}$$=$$410\textrm{
}\mu\textrm{C}/\textrm{m}^2$ and $A_{axis}$$=$$180\textrm{
}\mu\textrm{C}/\textrm{m}^2$ for BCGO. Using these parameters, the
field dependence of every component of the static
$\chi^{em}_{\delta\gamma}=\partial P_{\delta}/\partial H_{\gamma}$
ME tensor can be calculated for BCGO according to
Eqs.~\ref{Pa}-\ref{Pc}.

For these three compounds, several elements of the static ME tensor,
which are used in the present study for comparison with the
directional dichroism spectra, can be directly determined from the
measured field dependence of the ferroelectric polarization reported
in the literature. Only in those cases when experimental curves are
not available, the ME tensor elements are evaluated using the fitted
parameters as described above.

Fig.~\ref{fig1}(a) displays the ferroelectric polarization induced
along the [001] axis in BCGO by magnetic fields applied parallel to
the [110] direction, $P_{[001]}(H_{[110]})$, as reproduced from Ref.
\onlinecite{Murakawa}. The field dependence of the
$\chi^{em}_{[001],[110]}$ static ME tensor element for external
fields along the [110] axis, given by the derivative $\partial
P_{[001]}/\partial H_{[110]}$, is shown in Fig.~\ref{fig1}(c). Via
the ME sum rule in Eq.~\ref{KhiME_DC}, this element of the static ME
tensor is related to the integral of the directional dichroism
spectrum in the Voigt configuration, where the magnetic component of
light is parallel to the static magnetic field applied along the
[110] direction and the electric component of light is parallel to
the [001] axis. In this configuration, the directional dichroism
spectra reported for BCGO by Ref.~\onlinecite{Kezsmarki2011}
correspond to the difference of the red and blue curves in
Fig.~\ref{fig1}(b), which are the absorption spectra obtained for
counter-propagating THz waves. The comparison between the static and
optical data using Eq.~\ref{KhiME_DC} is shown Fig.~\ref{fig1}(c).

The following part of Fig.~\ref{fig1} shows similar analysis for
other three elements of the ME tensor in BCGO. In two cases, data
for SCSO and CCSO are also included. The dependence of the
ferroelectric polarization on the orientation of a constant field
$H$ is shown in panels (d), (g) and (j). The directional dichroism
spectra in these three cases are displayed in panels (e), (h) and
(k), while the comparison between the static and optical data is
given in panels (f), (i) and (l).

The $P_{[100]}(\theta)$ curve in Fig.~\ref{fig1}(g) is reproduced
from Ref.~\onlinecite{Murakawa}, where $\theta$ is the angle of the
magnetic field relative to the [001] axis. Since the tilting of the
magnetic field from the [010] direction by a small angle of
$\delta\theta$ introduces a weak transversal field $\delta{\bf
H}$$=$$(0,0,H\textrm{sin}\delta\theta)$, for ${\bf
H}$$\parallel$[010] $\chi^{em}_{[100][001]}$$=$$\partial
P_{[100]}/\partial H_{[001]}$$\approx$$1/H$$\times$$\partial
P_{[100]}/\partial \theta$. The corresponding optical experiment can
be realized in the Faraday configuration, where ${\bf
H}$$\parallel$[010], while the electric and magnetic components of
light are parallel to the [100] and [001] axes, respectively. These
THz absorption spectra are shown for the two opposite wave
propagation directions in Fig.~\ref{fig1}(h) as reproduced from Ref.
\onlinecite{Bordacs2012} for BCGO and from Ref.
\onlinecite{Ca2CoSi2O7} for CCSO.

The $P_{[001]}(\phi)$ curve in Fig.~\ref{fig1}(d) is taken from
Ref.~\onlinecite{Murakawa} and the $P_{[110]}(\theta)$ curve in
Fig.~\ref{fig1}(j) is calculated using Eqs.~\ref{Pa}-\ref{Pc}. In
the former and later cases, the elements of the static ME tensor are
respectively obtained according to
$\chi^{em}_{[001][010]}$$=$$P_{[001]}/\partial
H_{[010]}$$\approx$$1/H$$\times$$\partial P_{[001]}/\partial \phi$
for ${\bf H}$$\parallel$[100] and
$\chi^{em}_{[110][001]}$$=$$P_{[110]}/\partial
H_{[001]}$$\approx$$1/H$$\times$$\partial P_{[110]}/\partial \theta$
for ${\bf H}$$\parallel$[110]. The corresponding THz absorption
spectra are shown in panels (e) and (k), respectively.

\section{Discussion}
The comparison between the ME tensor elements calculated from the
static and optical data in the last column of Fig. \ref{fig1}
supports the applicability of the ME sum rule in these multiferroic
compounds. The magnitude and the field dependence of the static and
optical data in panels (f), (i) and (l) show quantitative agreement.
Their difference can be attributed to the following factors: i) the
directional dichroism measurements were performed at $T$$=$4\,K,
while the static experiments were carried out at $T$$\leq$2\,K where
the ME coefficients are larger by $\sim$10-20\%, ii) the two set of
experiments were performed on samples from different growths, iii)
in Fig.~\ref{fig1}(e) and (h) the polarization of light beams can
change during the propagation through the samples due to natural and
magnetic circular dichroism, iv) the model used to calculate the
field dependence of the static ME coefficients is not accurate due
to the linear field dependence of the magnetization assumed here to
reduce the number of fitting parameters, and v) uncertainty in the
geometrical factors of samples used in the static and optical
experiments may also cause an error of typically $\sim$10-20\%.

In the last three rows of Fig.~\ref{fig1} the dominant contribution
to the integral in the ME sum rule comes from the Goldstone mode,
whose excitation energy is proportional to the easy-plane component
of the static magnetic field. This mode has a very small gap of less
than $0.075$\,THz in zero field.\cite{Karlo_prl} In the field region
investigated here, its energy remains considerably smaller than
those of the other magnon modes. Hence, it dominates the integral in
the Eq.~\ref{KhiME_DC} sum rule due to the $\omega^2$ frequency
denominator. This mode is not allowed in an easy-plane magnet if the
magnetic component of light is parallel to the static magnetic field
as seen in Fig.~\ref{fig1}(b). Correspondingly, in
Fig.~\ref{fig1}(c) the ME tensor element calculated from the
directional dichroism data is smaller than those for the transverse
spin excitations shown in panels (f), (i) and (l).

Moreover, the ME tensor element calculated from the sum rule in
Fig.~\ref{fig1}(c) is one order of magnitude smaller than the value
determined from the static measurement, though they both change sign
in the same field region of $\mu_0H$$=$$7$$-$$9$\,T. This
significant difference may come from directional dichroism exhibited
by excitations out of range of our optical detection. Since all
magnon modes expected in the microscopic spin model of BCGO are
observed in the absorption experiments,\cite{Karlo_prl} we think
that low-energy phonon modes can show strong optical magnetoelectric
effect due to coupling to magnon modes. Though directional dichroism
has not been directly observed for phonon modes, recent optical
studies on multiferroic Ba$_3$NbFe$_3$Si$_2$O$_{14}$ reported about
the magnetoelectric nature of low-energy lattice
vibrations.\cite{Chaix2012} As another possibility, spin excitations
located out of our experimental window and not captured by the
spin-wave theory can also contribute to the directional dichroism
spectrum.

Besides the comparative analysis of static and optical ME data
carried out for the three compounds above to demonstrate the
applicability of the ME sum rule, we also make predictions for the
same and other multiferroic materials. Previous studies report about
magnetically induced ferroelectric polarization in the paramagnetic
phase of BCGO\cite{Murakawa2} and
SCSO\cite{Sr2CoSi2O7_lowtemp,Sr2CoSi2O7_prb} up to $T$$=$$300$\,K,
while their magnetic ordering temperature is $T_{N}$$\approx$$7$\,K.
Since the magnetic symmetry (MPG) of these compounds depends on the
orientation of the magnetic field but it is the same for the ordered
and the paramagnetic state, we expect that the directional dichroism
observed below $T_N$ in various configurations can survive up to
room temperature.

In the non-centrosymmetric soft magnet (Cu,Ni)B$_2$O$_4$ the
electric control of the magnetization direction has been
demonstrated together with directional dichroism of near-infrared
electronic excitations.\cite{Saito_Nmat} Since the contribution from
these d-d transitions to the ME sum rule is negligible due to their
high frequency and the $\omega^2$ denominator in Eq.~\ref{KhiME_DC},
we expect that directional dichroism should also be present for
low-frequency magnon excitations in this material.

The magnetic control of the ferroelectric polarization and/or the
electric control of the magnetization have been observed in a
plethora of multiferroic materials including perovskite manganites
with cycloidal spin order,\cite{KimuraNature, Murakawa_Eu, MbyE, PbyH} the room temperature
multiferroics BiFeO$_3$\cite{BiFeO_domain_control, BiFeO_Lee} and
Sr$_3$Co$_2$Fe$_{24}$O$_{41}$.\cite{Kimura2010, Chun2012} Based on the ME
sum rule, we predict that these compounds can also show directional
dichroism as already has been found for
Eu$_{0.55}$Y$_{0.45}$MnO$_{3}$\cite{EuYMnO} and
Gd$_{0.5}$Tb$_{0.5}$MnO$_3$\cite{Takahashi2013} in the spectral
range of the magnon excitations.

\section{Conclusions}
We derived a ME sum rule and discussed its validity for
non-centrosymmetric magnets. We showed that the ME sum rule can be
used to predict the static ME properties based on the directional
dichroism spectra governed by the optical ME effect and vica versa,
whenever the ME susceptibility of a material is antisymmetric with
respect to the time reversal. We verified this approach by a
quantitative comparison between static ME coefficients and
directional dichroism spectra experimentally determined for three
multiferroic compounds in the melilite family. In most cases we
found that the dominant contribution to the ME sum rule comes from
magnon excitations located in the GHz-THz region. Our approach is
applicable to most of the magnetoelectric multiferroics, where the
magnetically induced electric polarization can be controlled by the
magnitude or the direction of external magnetic field.\\

We thank Y. Tokura, H. Murakawa and K. Penc for valuable
discussions. This project was supported by Hungarian Research Funds
OTKA K108918, T\'AMOP-4.2.2.B-10/1-2010-0009,
T\'AMOP-4.2.1/B-09/1/KMR-2010-0002, T\'AMOP 4.2.4.
A/2-11-1-2012-0001, by the Estonian Ministry of Education and
Research under Grants SF0690029s09 and IUT23-03, by the Estonian
Science Foundation under Grants ETF8170 and ETF8703 and by the
bilateral program of the Estonian and Hungarian Academies of
Sciences under the Contract No. SNK-64/2013. S.B. was supported by
the Funding Program for World-Leading Innovative R\&D on Science and
Technology (FIRST Program), Japan.

\appendix*
\section{Derivation of the magnetoelectric sum rule from the Kubo formula}
The microscopic description of the linear response of a quantum
system to external stimuli is given by the Kubo formula. For the
frequency dependence of the ME susceptibility tensor, the
finite-temperature Kubo formula reads
\begin{equation}
\chi^{me}_{\gamma\delta}(z)=-\frac{1}{\hbar}\sum\limits_{m,n}\frac{e^{-\beta
\hbar\omega_{n}}-e^{-\beta
\hbar\omega_{m}}}{\sum\limits_{i}e^{-\beta
\hbar\omega_{i}}}\frac{\Bra{n}M_{\gamma}\Ket{m}\Bra{m}P_{\delta}\Ket{n}}{z-\omega_{m}+\omega_{n}},
\end{equation}
where $z$$=$$\omega$$+$$i\varepsilon$ and $\textrm{
}\varepsilon$\,$\xrightarrow{}$$0$$+$. $M_{\gamma}$ and $P_{\delta}$
are the magnetic and electric dipole operators, respectively.
$\Ket{m}$ and $\Ket{n}$ are eigenstates of the unperturbed system
with energies of $\hbar \omega_{m}$ and $\hbar\omega_{n}$, while
$\beta$ is the inverse temperature. In the zero-temperature limit
the Boltzmann factors vanish except for the $\Ket{0}$ zero energy
ground state:
\begin{equation}
\chi^{me}_{\gamma\delta}(z)=-\frac{1}{\hbar}\sum\limits_{m}\left(\frac{\Bra{0}M_{\gamma}\Ket{m}\Bra{m}P_{\delta}\Ket{0}}{z-\omega_{m}}-\frac{\Bra{0}P_{\delta}\Ket{m}\Bra{m}M_{\gamma}\Ket{0}}{z+\omega_{m}}\right).
\label{KhiME_Full}
\end{equation}
If  $\chi^{me}_{\gamma\delta}(\omega)$ is antisymmetric with respect
to the time reversal,
the $\Bra{0}M_{\gamma}\Ket{m}\Bra{m}P_{\delta}\Ket{0}$ product of
the transition matrix elements of the magnetic and electric dipole
operators is real.
The imaginary part of the transition matrix element product vanishes
since the magnetic dipole operator changes sign under time reversal
operation, which also requires the conjugation of the matrix
elements due to the exchange of the initial and final states.
The Kubo formula at zero temperature for the real and imaginary part
of the magnetoelectric susceptibility yields:
\begin{align}
\Re\chi^{me}_{\gamma\delta}(\omega)&=-\frac{2}{\hbar}\mathcal{P}\sum\limits_{m} \frac{\omega_{m}\Bra{0}M_{\gamma}\Ket{m}\Bra{m}P_{\delta}\Ket{0}}{\omega^2-\omega_{m}^2},\label{KhiME_Re}\\
\Im\chi^{me}_{\gamma\delta}(\omega>0)&=\frac{\pi}{\hbar}\sum\limits_{m}\Bra{0}M_{\gamma}\Ket{m}\Bra{m}P_{\delta}\Ket{0}\delta
(\omega-\omega_{m}). \label{KhiME_Im}
\end{align}
These expressions can also be obtained by second order perturbation
theory.\cite{Barron} With
$\Delta\alpha(\omega)=\frac{4\omega}{c}\Im\chi^{me}_{\gamma\delta}(\omega)$
one can reproduce  Eq.~\ref{KhiME_DC}:
\begin{align}
\chi^{me}_{\gamma\delta}(0)&=\frac{2}{\hbar}\mathcal{P}\sum\limits_{m}\frac{\Bra{0}M_{\gamma}\Ket{m}\Bra{m}P_{\delta}\Ket{0}}{\omega_{m}}\notag\\
&=\frac{2}{\hbar}\mathcal{P}\sum\limits_{m}\int^{\infty}_{0}{\frac{\Bra{0}M_{\gamma}\Ket{m}\Bra{m}P_{\delta}\Ket{0}}{\omega}}\cdot\delta{(\omega-\omega_{m})\textrm{d}\omega}\notag\\
&=\frac{c}{2\pi}\mathcal{P}\int^{\infty}_{0}{\frac{\Delta\alpha(\omega)}{\omega^{2}}\textrm{d}\omega}.
\label{KhiME_DC_Kubo}
\end{align}

\end{document}